\documentclass[acmsmall,nonacm]{acmart}

\usepackage{cleveref}
\usepackage{subcaption}
\usepackage{tabularray}
\UseTblrLibrary{booktabs}

\begin{document}

\title{Are You Really Empathic? Evidence from Trait, State and Speaker-Perceived Empathy, and Physiological Signals}

\author{Md Rakibul Hasan}
\affiliation{%
    \institution{Curtin University}
    \city{Bentley WA 6102}
    \country{Australia}%
}
\email{rakibul.hasan@curtin.edu.au}

\author{Md Zakir Hossain}
\affiliation{%
    \institution{Curtin University}
    \city{Bentley WA 6102}
    \country{Australia}%
}
\email{zakir.hossain1@curtin.edu.au}

\author{Aneesh Krishna}
\affiliation{%
    \institution{Curtin University}
    \city{Bentley WA 6102}
    \country{Australia}%
}
\email{a.krishna@curtin.edu.au}

\author{Shafin Rahman}
\affiliation{%
    \institution{North South University}
    \city{Dhaka 1229}
    \country{Bangladesh}%
}
\email{shafin.rahman@northsouth.edu}

\author{Tom Gedeon}
\affiliation{%
    \institution{Curtin University}
    \city{Bentley WA 6102}
    \country{Australia}%
}
\email{tom.gedeon@curtin.edu.au}


\begin{abstract}
When someone claims to be empathic, it does not necessarily mean they are \emph{perceived} as empathic by the person receiving it. Empathy promotes supportive communication, yet the relationship between listeners’ trait and state empathy and speakers’ perceptions remains unclear. We conducted an experiment in which speakers described a personal incident and one or more listeners responded naturally, as in everyday conversation. Afterwards, speakers reported perceived empathy, and listeners reported their trait and state empathy. Reliability of the scales was high (Cronbach’s $\alpha = 0.805$--$0.888$). Nonparametric Kruskal-Wallis tests showed that speakers paired with higher trait-empathy listeners reported greater perceived empathy, with large effect sizes. In contrast, state empathy did not reliably differentiate speaker outcomes. To complement self-reports, we collected electrodermal activity and heart rate from listeners during the conversations, which shows that high trait empathy listeners exhibited higher physiological variability.
\end{abstract}

\ccsdesc[500]{Human-centered computing~Empirical studies in HCI}
\ccsdesc[500]{Applied computing~Psychology}
\ccsdesc[300]{Applied computing~Health informatics}

\keywords{Empathy, State, Trait, Perceived, Physiological Signal, Electrodermal Activity, Heart Rate}

\begin{teaserfigure}
    \centering
  \includegraphics[width=0.8\textwidth]{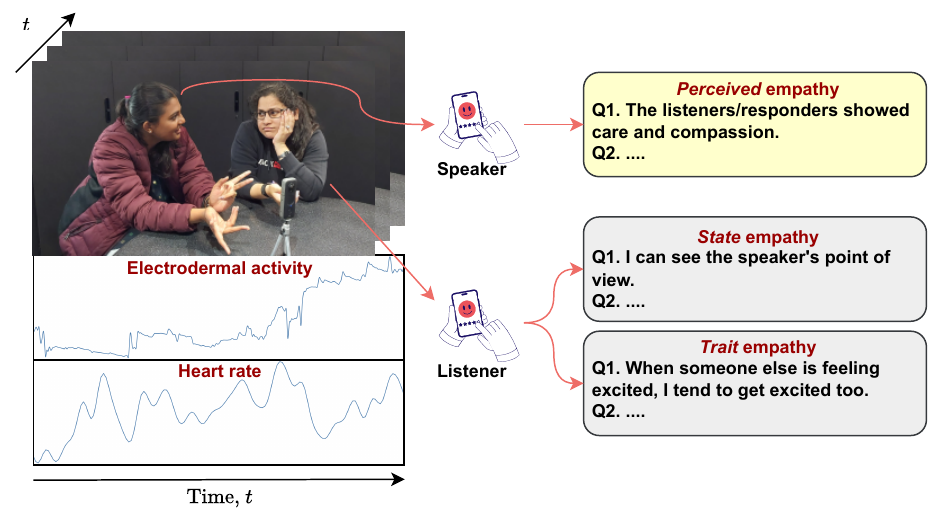}
  \caption{\textbf{Study design overview}: Speakers narrated a personal incident while listeners responded naturally. Speakers rated perceived empathy, while listeners reported state and trait empathy. Concurrently, listeners’ physiological responses (electrodermal activity, heart rate) were recorded using a wearable device.}
  \Description{The figure shows a speaker and listener in conversation, with physiological signals (electrodermal activity and heart rate) recorded from the listener. On the right, examples of questionnaire items illustrate three constructs: perceived empathy from the speaker’s perspective (``The listeners/responders showed care and compassion''), state empathy from the listener’s perspective (``I can see the speaker’s point of view''), and trait empathy from the listener’s perspective (``When someone else is feeling excited, I tend to get excited too'').}
  \label{fig:teaser}
\end{teaserfigure}


\maketitle

\section{Introduction}
Empathy is largely a spontaneous, vicarious reaction to the situation and perspective of someone else \citep{hoffman1978toward}. Central to supportive communication and to the design of humane interactive systems, it shapes trust, engagement and help-seeking in mediated contexts \citep{genc2024situating}. With its prominence in discourse, many studies investigated how empathy is expressed or detected in various content types, including text, audiovisual, audio and physiological signals \citep{hasan2025empathy}.

Empathy is often sub-categorised into multiple facets, such as cognitive (understanding someone's perspective) vs affective (responding with appropriate emotion) empathy \citep{cuff2016empathy} and state (situational) vs trait (fixed) empathy \citep{shen2023modeling}. State empathy refers to someone's momentary empathy towards a particular stimulus in a particular situation, whereas trait empathy refers to someone's fixed general empathic capabilities \citep{shen2023modeling}. Among different kinds of empathy, distinguishing between cognitive and affective empathy has been primarily studied \citep{hasan2025empathy,dey2022enriching}. Although Psychology literature \citep{heyers2025state} distinguishes between trait and state empathy, their differential impact on perceived empathy in dialogue remains underexplored. In parallel, HCI and natural language processing literature explored measuring perceived empathy in conversations and dialogue systems \citep{xu2024multi,concannon2024measuring}. However, empirical human-human studies linking trait and state empathy of listeners to speakers’ perceptions in unscripted talk are scarce.

Physiological signals offer complementary windows on empathic engagement. Prior work reports Electrodermal Activity (EDA) and Heart Rate (HR) changes during empathic communication and links interpersonal synchrony to rapport and empathic processes \citep{finset2011electrodermal,stuldreher2020physiological,qaiser2023shared}. HCI has also shown that exposing or leveraging physiological signals can shape empathy in mediated interaction \citep{winters2021can}. Most studies, however, rely on controlled stimuli or specialised labs, limiting ecological validity.

Understanding and measuring empathy has gained traction over recent years, predominantly on textual content \citep{hasan2025empathy}. A few studies explored the potential of physiological signals. For example, \citet{golbabaei2022physiological} analysed electrocardiogram and skin conductance for empathy of people having autism spectrum disorder. \citet{wei2021effective} studied empathy of cocaine-dependent subjects through resting-state functional magnetic resonance imaging. \citet{kuijt2020prediction} uses EEG cortical asymmetry of different areas of the brain to predict someone's empathy. Contrary to these studies, we use EDA and HR, which are easy to measure through wearable devices.

We address \emph{how a listener’s empathic trait or momentary state maps onto a speaker’s perceived empathy during natural conversation}. \Cref{fig:teaser} illustrates our study design: speakers described a personal incident while listeners responded naturally, listeners’ physiological signals were recorded, and both parties completed questionnaires capturing perceived, state and trait empathy. During the conversation, we record EDA and HR from wearable devices on the listeners. We test whether listener trait or state empathy better predicts speakers’ perceived empathy, and we examine whether simple EDA/HR features track empathic disposition. Our primary contributions include:
\begin{enumerate}
    \item Evidence that listener \emph{trait} empathy, more than \emph{state} empathy, predicts speakers’ perceived empathy in natural conversation.
    \item Physiological analyses showing that higher trait empathy coincides with higher variability. 
    \item A practical multimodal protocol for studying empathy in unscripted natural conversations that includes validated questionnaires and wearable sensing, which would help design and evaluation of empathic technologies.
\end{enumerate}


\section{Method}
Our experiment captures how listeners’ empathy influences speakers’ experiences during emotionally charged interactions. In each session, one participant assumed the role of speaker and shared a personally significant concern or experience, while others (one or multiple) acted as listeners and responded naturally. This unscripted format was chosen to capture the richness of real-life empathic interactions, in contrast to scripted or role-played settings that may constrain authentic behaviour. The experiment includes subjective self-report measures with objective physiological recordings to provide a multimodal assessment of empathy.


\subsection{Participant} 
Participants were recruited from university students and staff, primarily through classroom announcements and further through convenience sampling \citep{baltes2022sampling}. The inclusion criteria required participants to have normal (or corrected to normal) hearing ability, normal English speaking and understanding ability, and to be comfortable discussing personal experiences in a research setting. Participants received no compensation for their time. Participants themselves selected their roles (speaker or listener) before each experiment. A total of 34 speaker-listener pairs from 18 unique participants took part in this study. Some participants took part both as listeners and as speakers in different sessions, which made the total speaker count 23 and the listener count 34, with 14 unique speakers and 15 unique listeners. We considered this sample size appropriate according to the \emph{local standard} of sample size selection \citep{caine2016local}.

Participants were aged 20 to 35 years old, with a mean age of 25.2 years and a standard deviation of 4.5. There were eleven male and six female participants. Ten of them were Bachelor's or Honours students, three were postgraduate students, and two were research staff. All were fluent English speakers, with eight participants speaking English as their native language. In terms of ethnicity, there were nine Asian, seven Caucasian and one Afghan. Among all participants, three identified themselves as neurodiverse (e.g., autism spectrum disorder or attention deficit hyperactivity disorder). One participant did not volunteer their demographic information.

\subsection{Procedure}
All procedures were approved by the Human Research Ethics Office of the host university (Approval No. HRE2024-0613). Each experimental session followed a standard protocol, including the following major steps:
\begin{enumerate}
    \item \textbf{Preparation}: Participants were seated side-by-side, slightly facing one another, in a quiet room (see \Cref{fig:teaser}). A Participant Information Form and a Consent Form were served and demonstrated to the participants. The speaker was instructed to share a recent emotionally meaningful experience (either positive or negative), choosing from a list of 27 suggested topics (e.g., academic stress, relationship issues or celebrations) or one of their own \citep{hasan2024thesis}. They discussed among themselves to decide topics and their roles. The speakers decided on topics in consultation with the listeners to make sure everyone was comfortable discussing the chosen topic.
    \item \textbf{Conversation}: The listener was instructed to respond naturally, without specific guidance on how to display empathy. Participants were asked to aim for 2--3 minutes, but no time limit or reminders were imposed; conversations continued as long as needed. Throughout the conversation, we collected physiological signals from the listeners using a wearable device.
    \item \textbf{Post-conversation questionnaires}: Both speaker and listener completed their respective self-report instruments via Qualtrics.
\end{enumerate}

\subsection{Instruments}
\subsubsection{Self-Reported Assessments}
Following the interaction, all participants (both speaker and listener) fill in a questionnaire (Appendix \Cref{tab:ques}). The questionnaire includes four sections: (1) 6 demographic questions; (2) 13 questions for speakers, where 2 questions are about their expressed emotion, and the remaining 11 are about their \emph{perceived} empathy and assessment of listeners from their point of view; (3) 10 questions for listeners about their expressed emotion and empathy (\emph{state} empathy), including 1 question about how long the listener personally know the speaker; and (4) 16 questions for listeners about their \emph{trait} empathy. Except for the demographic questions, answers for all other questions were in terms of an 11-point Likert scale, with 0 being ``Not at all``, 10 being ``Completely agree'' and 5 being a neutral opinion in case they are unsure. Several of these questions are adapted from the literature: seven of the speaker questions from the Consultation and Relational Empathy (CARE) scale \citep{mercer2004consultation}, eight state empathy questions from prior experimental empathy studies \citep{mathur2021modeling,shen2010on}, and all 16 trait empathy questions from the Toronto Empathy Questionnaire (TEQ) \citep{spreng2009toronto}. Trait empathy questions were administered only once per listener, because, by definition, it is a stable individual characteristic rather than a state that varies across sessions.

\subsubsection{Physiological Signals}
Listeners’ physiological responses were recorded using an EmotiBit MD\footnote{\url{https://www.emotibit.com/product/emotibit/}} wearable device on the wrist of their non-dominant hand. EmotiBit collected several physiological signals. Among these, we primarily analyse EDA and HR (HR) since prior research has demonstrated their connection to emotion and empathy research \citep{tapus2008socially,hossain2017classifying, egger2019emotion}.

\subsection{Data Analysis}
While null-hypothesis significance testing with $p$-values has long been the dominant approach, it has well-documented flaws \citep{ho2019moving}. $p$-values indicate whether a statistical difference exists but not the \emph{magnitude} of that difference. Effect sizes, in contrast, quantify the strength of the association and capture practical significance \citep{mahmud2023which}. Moreover, $p$-values are highly dependent on sample size \citep{nakagawa2007effect}, whereas effect sizes are not. For these reasons, several authors recommend replacing or at least supplementing $p$-values with effect size estimates \citep{cumming2014the,halsey2015the}. Guided by these recommendations, we primarily use effect sizes when comparing variables.

\subsubsection{Self-Reported Questionnaire}
Some items in the questionnaires are worded negatively, for example ``Other people’s misfortunes do not disturb me a great deal.'' in the TEQ. To ensure that higher Likert scores consistently reflect greater empathy, we reverse-coded these items before computing the scale totals.

Given the ordinal structure of Likert-type scales, we employed nonparametric methods that do not assume normal distribution. Specifically, we used the Kruskal-Wallis $H$ test \citep{Kruskal1952use} to evaluate group-level differences, such as comparing self-reported outcomes across empathy quartiles. This test is a rank-based analogue of one-way ANOVA and is widely recommended for ordinal data \citep{mckight2010kruskal}.

\paragraph{Effect Size and Confidence Intervals.}
For each Kruskal-Wallis test, we report $\eta^2$ as an effect size \citep{tomczak2014the}:
\begin{equation}
\eta^2 = \frac{H - k + 1}{N - k},
\end{equation}
where $H$ is the Kruskal-Wallis statistic, $k$ is the number of groups and $N$ is the total sample size. 

The best practices for reporting effect sizes emphasise providing confidence intervals alongside point estimates \citep{vacha2004estimate}. To obtain 95\% confidence intervals for $\eta^2$, we applied a stratified percentile bootstrap procedure. Within each group $i$, we resampled $n_i$ observations with replacement, recomputed the Kruskal-Wallis statistic $H$ and the effect size $\eta^2$ and repeated this procedure 500 times. The empirical distribution of bootstrapped $\eta^2$ values was then used to derive the 2.5th and 97.5th percentiles as the confidence bounds.

\subsubsection{Physiological Signals}
Physiological signals often contain noise and baseline drifts that need to be addressed prior to analysis. Following \citep{xia2015active}'s approach on EDA signal preprocessing, we applied a 6th-order low-pass Butterworth filter with a cutoff frequency of 0.5~Hz for EDA and 0.1~Hz for HR signals. Signals were standardised within each participant by $z$-scoring (subtracting the mean and dividing by the standard deviation of that participant’s time series). This normalisation step controlled for inter-individual differences in baseline physiology and emphasised within-session variation. 

From the preprocessed EDA and HR time series, we extracted four summary statistics for each participant: mean, median, standard deviation and lag-1 autocorrelation. The reason behind choosing these four summary statistics is two-fold. First, prior works on empathy \citep{mathur2021modeling,hasan2025tfmpathy} also used these four statistics to extract a fixed feature set from time series data. Second, the mean and median capture the \emph{tonic} component, which is the slower baseline variations associated with general arousal levels. The standard deviation and autocorrelation capture the \emph{phasic} component, the rapid, transient reactions to discrete stimuli or emotional instances \citep{boucsein2012electrodermal}. Overall, the mean and median capture the central tendency, the standard deviation reflects variability, and the autocorrelation captures temporal structure.

To examine whether empathic tendencies were reflected in physiological signals, we compared listeners in high versus low empathy groups (median split, as in \citep{mathur2021modeling}).




\section{Results \& Discussion}

\subsection{Descriptive Statistics}
Speakers were asked to indicate the primary emotion they expressed during the conversation. The most frequently reported emotions were \emph{fear}, \emph{disappointment} and \emph{sadness}, with smaller counts across other categories. This indicates that most conversations are centred on negatively valenced experiences. Speakers also rated the intensity of their expressed emotion on an 11-point scale. The average intensity was 8.2 (SD = 1.6), suggesting that participants generally shared emotionally charged incidents.

The average perceived empathy score was 72.8 (SD = 16.9) out of a maximum of 99, indicating moderately high perceived empathic responses. The average score of listeners' \emph{state} empathy was 75.1 (SD = 14.0) out of 99, and \emph{trait} empathy was 111.7 (SD = 20.6) out of 176. All scores were computed as the sum of the respective items for each scale.

\subsection{Reliability of Empathy Scales}

We first assessed internal consistency of the scales using Cronbach’s $\alpha$ \citep{cronbach1951coefficient}. As shown in \Cref{tab:reliability}, all three scales demonstrated high reliability \citep{tavakol2011making}, with $\alpha$ varying between $0.805$ and $0.888$.

\begin{table}[t!]
    \centering
    \caption{Cronbach’s $\alpha$ \citep{cronbach1951coefficient} for the three empathy scales used in this study.} \label{tab:reliability}
    \begin{tabular}{lcc}\toprule
    \textbf{Empathy type} & $\mathbf{\alpha}$ & \textbf{95\% CI} \\\midrule
    Speaker perceived empathy & 0.870 & [0.792, 0.926] \\
    Listener state empathy & 0.888 & [0.822, 0.937] \\
    Listener trait empathy & 0.805 & [0.695, 0.889] \\\bottomrule
    \end{tabular}
\end{table}

\subsection{Listener's Trait and State Empathy vs Speaker's Perceived Empathy}

We next examined whether speaker-reported outcomes varied across listener empathy groups, using Kruskal-Wallis tests with $\eta^2$ as the effect size estimate. 

\begin{figure}[t!]
    \centering
    \includegraphics[width=1.0\linewidth]{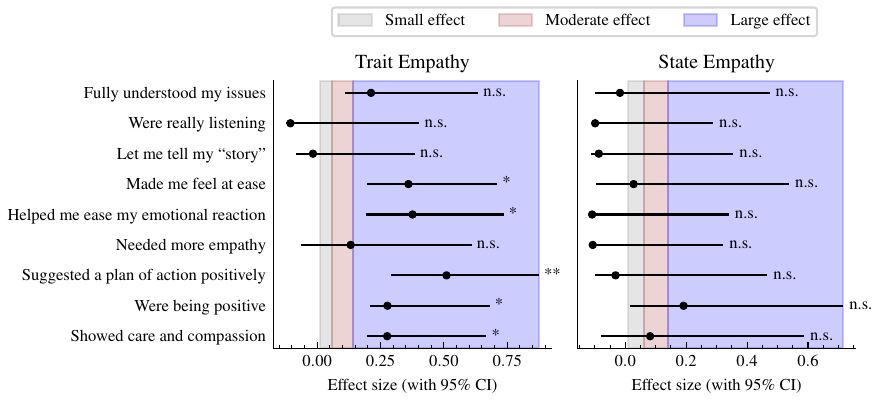}
    \caption{Effect sizes ($\eta^2$) with 95\% confidence intervals from Kruskal-Wallis tests examining whether speakers’ perceived empathy differed according to listeners’ trait empathy (left) or state empathy (right). Items correspond to individual speaker-reported questions. Vertical shaded bands indicate conventional benchmarks for small (grey, $\eta^2 \geq 0.01$), moderate (brown, $\eta^2 \geq 0.06$), and large (blue, $\eta^2 \geq 0.14$) effects \citep{kassambara-kruskal}. Asterisks denote statistical significance levels from the Kruskal-Wallis tests (* = $p<.05$; ** = $p<.01$, n.s. = not significant). Error bars represent bootstrap 95\% confidence intervals based on resampling within groups.}
    \label{fig:effect-size}
    \Description{Two side-by-side forest plots summarise Kruskal-Wallis effect sizes ($\eta^2$) with 95\% confidence intervals for speaker-reported items. Left panel: Trait Empathy; right panel: State Empathy. The y-axis lists nine items (e.g., ``Fully understood my issues'', ``Helped me ease my emotional reaction'', ``Showed care and compassion''). The x-axis shows $\eta^2$; shaded vertical bands mark conventional small, moderate and large effect size benchmarks. Points with horizontal error bars indicate estimates and confidence intervals; asterisks mark statistical significance. Several trait-based items are significant with larger effects, whereas state-based items show overlapping confidence intervals and no statistical significance.}
\end{figure}

\Cref{fig:effect-size} shows that speakers’ perceptions of empathy are more consistently associated with listeners’ trait empathy than with their state empathy. In the trait condition, several items show large effect sizes ($\eta^2 \geq 0.14$) with relatively narrow confidence intervals: ``Made me feel at ease'', ``Helped me ease my emotional reaction'', ``Suggested a plan of action positively'', ``Were being positive'', and ``Showed care and compassion''. These items also reach statistical significance, assuming a significance level of $0.05$. Both the effect size and the $p$-value indicate robust differences across trait-empathy groups. In contrast, state empathy effects were uniformly small, and their confidence intervals frequently crossed zero. None of the state-empathy comparisons reached statistical significance. These results indicate that trait empathy, as a stable individual characteristic, plays a more decisive role in shaping how speakers perceive their listeners’ empathy than transient, situational state empathy.

Notably, the strongest trait-based associations corresponded to supportive and relational behaviours (e.g., fostering ease, compassion, and positivity) rather than to more instrumental aspects such as ``Fully understood my issues'' or ``Were really listening'', which were not significant. This pattern suggests that dispositional empathy may be most visible to speakers when it manifests through overtly caring and emotionally responsive behaviours.

\subsection{Do High- vs Low Empathy Groups Differ in Physiological Features?}

\begin{figure}[t!]
  \centering
  \subcaptionbox{Trait Empathy}[1\textwidth]{\includegraphics[width=\linewidth]{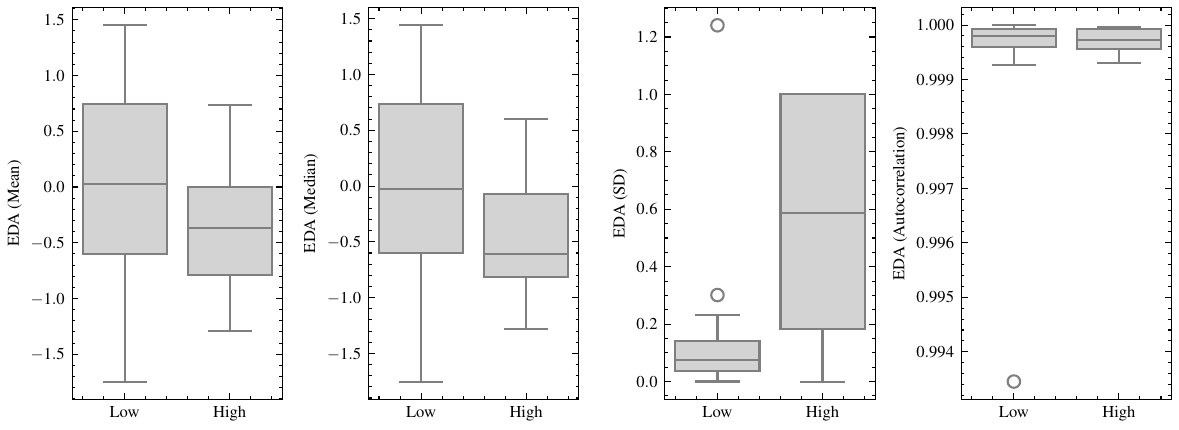}}
  \subcaptionbox{State Empathy}[1\textwidth]{\includegraphics[width=\linewidth]{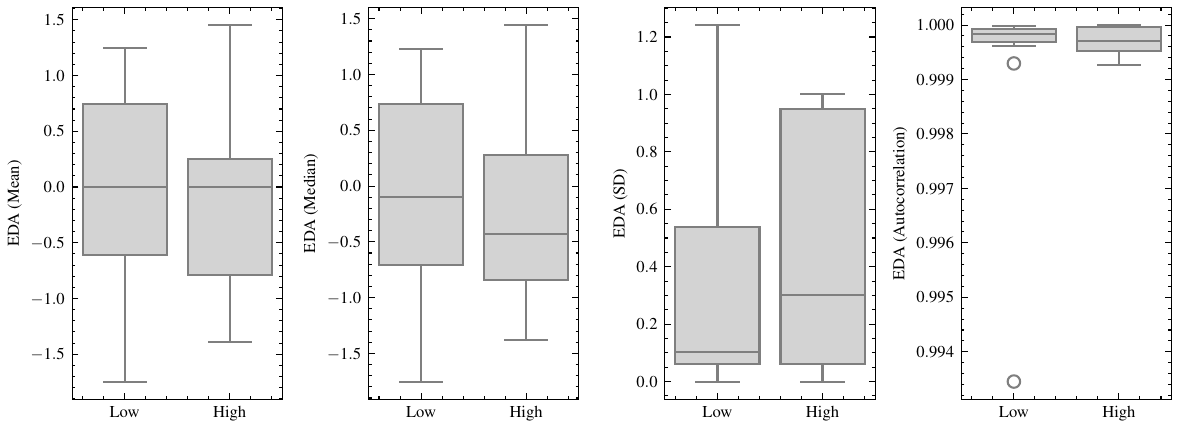}}
  \caption{Electrodermal activity (EDA) features (mean, median, standard deviation and autocorrelation) of listeners grouped by low vs high trait empathy (top row) and state empathy (bottom row). Values are z-scored within participants before aggregation.}\label{fig:eda}
  \Description{Eight boxplots display electrodermal activity features for low versus high empathy groups. Top row (trait Empathy): mean, median, standard deviation and autocorrelation; bottom row (state Empathy): the same four features. Each panel has two boxes labelled Low and High; values are $z$-scored within participants. Trait grouping shows higher variability (standard deviation) in the high empathy group; the mean and median differ little. State grouping shows broad overlap across features.}
\end{figure}

\begin{figure}[t!]
  \centering
  \subcaptionbox{Trait Empathy}[1\textwidth]{\includegraphics[width=\linewidth]{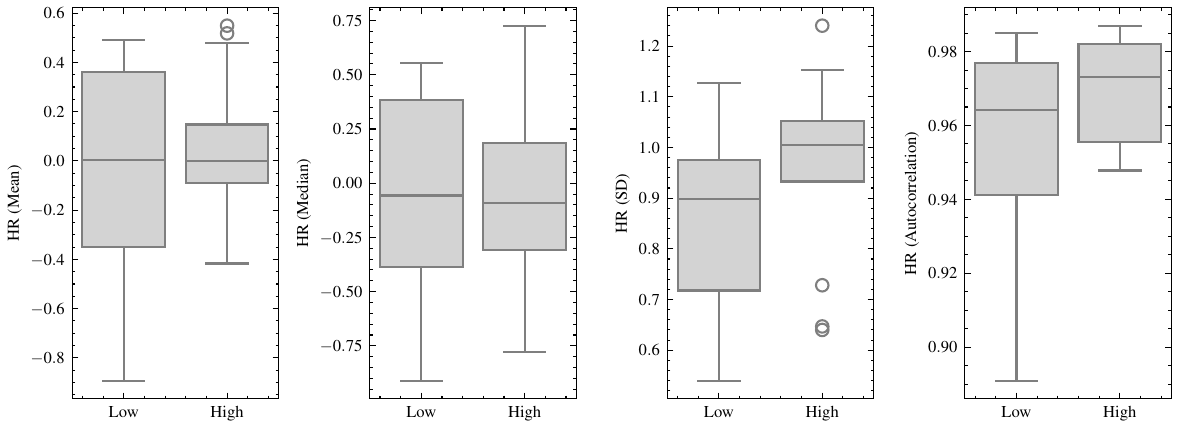}}
  \subcaptionbox{State Empathy}[1\textwidth]{\includegraphics[width=\linewidth]{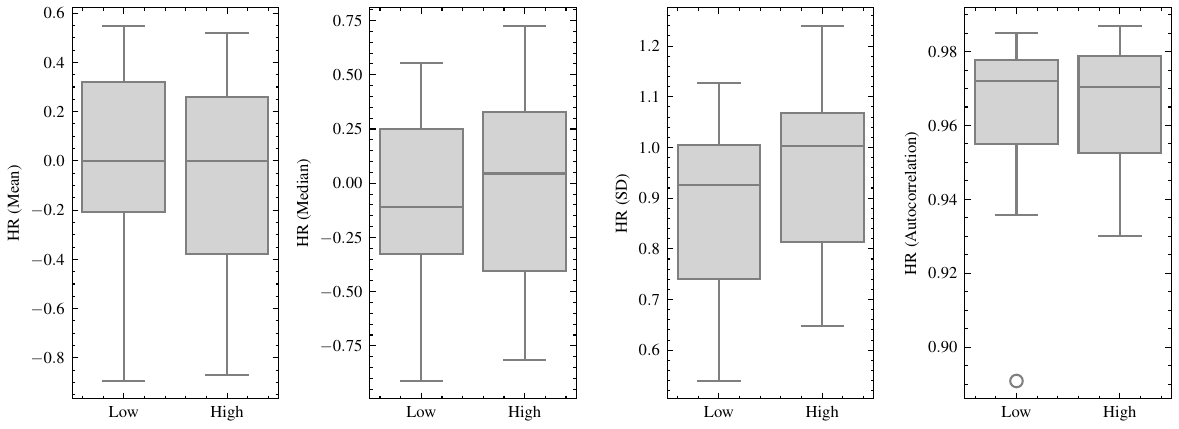}}
  \caption{Heart rate (HR) features (mean, median, standard deviation and autocorrelation) of listeners grouped by low vs high trait empathy (top row) and state empathy (bottom row). Values are z-scored within participants before aggregation.}\label{fig:hr}
  \Description{Eight boxplots display heart rate features for low versus high empathy groups. Top row (trait Empathy): mean, median, standard deviation and autocorrelation; bottom row (state Empathy): the same four features. Each panel has two boxes labelled Low and High; values are $z$-scored within participants. Trait grouping shows somewhat higher variability (standard deviation) and higher autocorrelation in the High group; the mean and median are similar. State grouping shows overlapping distributions across all features.}
\end{figure}

\Cref{fig:eda,fig:hr} present EDA and HR features grouped by low versus high empathy (trait and state). Values were $z$-scored within participants before aggregation.

For EDA (\Cref{fig:eda}), high-trait empathy listeners showed higher variability, as indicated by higher standard deviations compared to low-trait listeners. Mean and median EDA values showed small differences. By contrast, grouping by state empathy revealed no reliable separation: distributions overlapped substantially across all features.

HR features (\Cref{fig:hr}) showed a comparable pattern. High-trait empathy listeners had higher variability and slightly higher autocorrelation. Group differences in mean and median HR were small. As with EDA, state empathy groups showed broad overlap, with no consistent divergence across features.

Between trait and state empathy, both EDA and HR features suggest that trait empathy is more distinguishable than state empathy using physiological signals. The results also suggest that among the four summary statistics, standard deviation is most effective in distinguishing high versus low empathy groups.

\subsection{Limitation}
The present findings should be interpreted in light of some limitations. First, the sample size was modest, which constrains the generalisability of effects. We mitigated this by emphasising effect sizes, which are less sensitive to sample size than significance tests. Second, participants were drawn from a university population, yielding a relatively homogeneous group in terms of age and education. Replication with larger and more diverse samples may help confirm and extend these results.

\section{Conclusion}
This study examined how listeners’ trait and state empathy relate to speakers’ perceived empathy, including the perception of being understood and supported, as well as to listeners’ physiological responses during conversation. Across multiple self-report measures, trait empathy emerged as a stronger predictor of perceived empathy than state empathy, with significant and large effects for items reflecting care, compassion and relational warmth. Physiological analyses further suggested that high-trait empathy listeners display higher EDA and HR variability and higher HR autocorrelation. These results highlight that stable empathic traits, rather than transient situational states, may play a decisive role in shaping how empathy is communicated and experienced in interpersonal settings. From an HCI perspective, this implies that trait-level empathy indicators could be especially valuable for informing the design of empathic interfaces and digital support systems. Characterising both perceived and embodied empathy will enable the development of technologies that promote authentic, supportive human-computer interactions.
 

\bibliographystyle{ACM-Reference-Format}
\bibliography{refs}

\appendix

\section{Questionnaires}
\Cref{tab:ques} presents the questions participants (speakers and listeners) answered in this study.

\begin{table}[t!]
    \centering
    \caption{The questionnaires used to collect participants' demographic information and to assess the speaker's perceived empathy and the listener's state and trait empathy.}
    \label{tab:ques}
    \begin{tblr}{width=\textwidth, colspec={cX[l]}} \toprule
        SL & Question \\ \midrule
        \SetCell[c=2]{l} \textit{Demographics} & \\
        1 & How old are you? Please enter in years, for example, 19 \\
        2 & By what gender do you identify? \\
        3 & In which program are you currently enrolled? \\
        4 & What is your first or native language? \\
        5 & What is your ethnicity? \\
        6 & Do you identify as neurodiverse, e.g., autism spectrum disorder or attention deficit hyperactivity disorder? \\ \midrule
        \SetCell[c=2]{l} \textit{Speaker} & \\
        1 & Which is the main emotion you expressed? \\
        2 & What was the intensity of your expressed emotion? \\
        3 & I needed more empathy than what was shown by my peers during the conversation. \\
        4 & I needed more empathy than what was shown by my peers during the conversation. \\
        5 & The listeners'/responders' empathic behaviour helped me ease my emotional reaction. \\
        6 & The listeners/responders made me feel at ease. \\
        7 & The listeners/responders let me tell my ``story''. \\
        8 & The listeners/responders were really listening. \\
        9 & The listeners/responders fully understood my issues. \\
        10 & The listeners/responders showed care and compassion. \\
        11 & The listeners/responders were being positive. \\
        13 & The listeners/responders suggested a plan of action positively. \\ \midrule
        \SetCell[c=2]{l} \textit{Listener -- State Empathy} & \\
        1 & How long have you known the speaker? \\
        2 & The speaker’s emotions were genuine. \\
        3 & I experienced the same emotions as the speaker while having the conversation. \\
        4 & I could feel the speaker's emotions. \\
        5 & I can see the speaker's point of view. \\
        6 & I can understand what the speaker was going through. \\
        7 & The speaker's reactions to the situation were understandable. \\
        8 & I was fully absorbed in the conversation. \\
        9 & I can relate to what the speaker was going through. \\
        10 & I felt compassion towards the speaker’s situation. \\ \midrule
        \SetCell[c=2]{l} \textit{Listener -- Trait Empathy} & \\
        1--16 & Toronto Empathy Questionnaire (TEQ) \citep{spreng2009toronto} \\ \bottomrule
    \end{tblr}
\end{table}

\end{document}